\documentstyle[11pt]{elsart}
\begin{document}
\begin{frontmatter}

\title{Magnetization in quasiperiodic magnetic multilayers with biquadratic
exchange and uniaxial anisotropy}

\author{C.G. Bezerra and M.G. Cottam}
%\thanks[email]{Corresponding author, e-mail: ela@dfte.ufrn.br, Fax: +-55-84-
%2153791}

\address{Department of Physics and Astronomy,
University of Western Ontario, N6A 3K7, London, Ontario, Canada.}

\begin{abstract}
A theoretical study is made of the magnetization versus applied
field curves of ferromagnetic/nonmagnetic multilayers constructed
according to a Fibonacci quasiperiodic sequence. The ferromagnetic
films are assumed to have uniaxial anisotropy and are coupled by
both bilinear and biquadratic effective exchange. The effects of
quasiperiodicity in the magnetic phases are illustrated
numerically for Fe/Cr systems.



\end{abstract}
\end{frontmatter}

Magnetic multilayers have attracted attention due to their
physical properties and potential for technological applications
\cite{1}. The coupling between the magnetic layers often
corresponds to a bilinear (BL), or Heisenberg, exchange type.
However, some multilayers also exhibit a biquadratic (BQ) exchange
coupling, which favours a $90^\circ$ orientation between the
adjacent ferromagnetic layers \cite{2,3}. The BQ exchange,
although often weaker than the BL exchange, can be large in some
systems. This leads to new possibilities for the phase diagrams
and magnetization curves.

Furthermore, it is possible to build magnetic multilayers where
the film thickness and stacking are well controlled, as in
periodic superlattices. However, there has also been interest in
magnetic systems formed with the layers following a quasiperiodic
sequence. Recent studies, applied to Fe/Cr Fibonacci multilayers,
showed that a self-similar behaviour of the magnetic properties
occurs only when the BQ and BL exchange are comparable \cite{4,5}.
The authors included effects of a cubic anisotropy (see, e.g.,
\cite{6}). However, some magnetic multilayers may instead exhibit
uniaxial anisotropy \cite{7}, and this gives a new symmetry to the
system that, together with the BQ exchange and quasiperiodicity,
may lead to new configurations. This has motivated the present
work.

A Fibonacci structure is formed by juxtaposing two basic building
blocks, $A$ and $B$, following a Fibonacci sequence. For the
present work we choose ferromagnetic Fe and nonmagnetic Cr to
correspond to $A$ and $B$, respectively. By definition, the $Nth$
Fibonacci generation, $S_N$, is obtained by appending generation
$N-2$ to generation $N-1$, i.e., $S_N = S_{N-1} S_{N-2}$ taking
$S_0 = B$ and $S_1 = A$. Thus, for example, $S_3$ and $S_5$
correspond to the magnetic multilayers Fe/Cr/Fe and
Fe/Cr/Fe/Fe/Cr/Fe/Cr/Fe respectively.

We assume the magnetization of the ferromagnetic layers to be in
the $xy$-plane and take the $z$-axis as the growth direction.
Following \cite{4,5} but with the anisotropy replaced by a
uniaxial term as in \cite{7}, the magnetic energy per unit area is

\begin{eqnarray}
\lefteqn {{{E_{T}} \over
{tM_{S}}}=\sum_{i=1}^n(t_{i}/t){\{-H_{0}\cos(\theta_{i}-\theta_{H})-
{1 \over 2}H_{ua}{{\cos}^2{(\theta_{i}-\theta_{ua})}}\}}}
\nonumber
\\ && +\sum_{i=1}^{n-1}{\{-H_{bl}\cos(\theta_{i}-\theta_{i+1})
+H_{bq}{{\cos}^2{ (\theta_{i}-\theta_{i+1})}}\}}.
\end{eqnarray}

Index $i (= 1,2,\dots,n)$ labels the Fe layers in the $Nth$
generation structure. For example, $n = 4$ in the case of $N = 5$,
with three Fe layers having thickness $t$ and one having thickness
$2t$, where $t$ denotes the thickness of a single $A$ layer (the
basic tile). In layer $i$ (with thickness $t_i$) the equilibrium
magnetization is in the $xy$-plane at angle $\theta_i$ to the
$x$-axis. The first term in eq. (1) is the Zeeman energy due to an
applied field $H_0$ at an angle $\theta_H$. The second term is due
to a uniaxial anisotropy $H_{ua}$ with easy axis specified by
angle $\theta_{ua}$. The third and fourth terms (involving
parameters $H_{bl}$, $H_{bq}$) describe the BL and BQ exchange
between adjacent layers. The set ${\{\theta\}}_i$ of equilibrium
angles is calculated numerically by minimizing the magnetic energy
given by eq. (1). For that purpose we use two methods, namely, the
simulated annealing and the gradient methods (see \cite{5} for
descriptions). The simulated annealing method is based on the fact
that heating and then cooling a material slowly brings it into a
more uniform state, which is the minimum energy state. In this
process, the role played by a pseudo temperature $T$ is to allow
the configurations to reach higher energy states with probability
$p$ given by the Boltzmann law $p = \exp(- \Delta E/kT)$, where
$\Delta E$ is the energy difference. Energy barriers that would
otherwise force the configurations into local minima can then be
overcome. On the other hand, the gradient method is based on
finding the directional derivative of the magnetic energy in the
search for its global minimum in the n-dimensional space composed
of the variables ${\{\theta\}}_i$. It is the gradient of the
magnetic energy with relation to the angles that furnishing the
direction, and eventually the location, of the required global
minimum. Both cited methods are used for each value of the applied
magnetic field and for each set of magnetic parameters. We choose
the configuration with the lowest energy furnished by both methods
as giving the equilibrium configuration ${\{\theta\}}_i$. From the
calculated ${\{\theta\}}_i$, we obtain normalized values of the
net magnetization in the $H_0$ direction, appropriate to the Fe/Cr
system.

Magnetization curves found for $N = 3, 5$ and $7$ are shown in
Fig.\ 1. Here the in-plane magnetic field is applied along the
easy axis. The magnetic parameters are such that $H_{bq} <<
|H_{bl}|$ and $H_{bq} << H_{ua}$. For $N = 3$ (see Fig.\ 1a) the
magnetizations in adjacent layers are antiparallel at low fields.
As $H_0$ increases, a first-order phase transition occurs (at
$\sim 0.8$ kOe) to a spin-flop phase. Saturation is reached at
$H_0 \sim 1.9$ kOe. For $N = 5$ (see Fig.\ 1b), due to the
different thickness of the Fe layers, the net magnetization at low
fields has about $20\%$ of its saturation value. There are
first-order phase transitions at $H_0 \sim 1.0$ kOe and $H_0 \sim
1.6$ kOe. Saturation is reached at $\sim$ $2.9$ kOe. When $N = 7$
(Fig.\ 1c) there is again a non-zero magnetization at low fields.
There are three first-order phase transitions at $H_0 \sim 0.5$
kOe, $1.0$ kOe and $1.6$ kOe.

Another set of magnetization curves is shown in Fig.\ 2, taking a
larger BQ exchange and other parameters as before. For $N = 3$,
due to the strong $H_{bq}$ the adjacent layer magnetizations are
only approximately antiparallel at low field (corresponding to an
asymmetric phase \cite{7}). A first-order phase transition occurs
at $H_0 \sim 0.4$ kOe to a spin-flop phase and saturation is
reached at $H_0 \sim 4.3$ kOe. For $N = 5$ all transitions appear
continuous (second-order phase transitions) and saturation is
reached at $H_0 \sim 6.4$ kOe. For $N = 7$ there appears to be a
first-order phase transition at $H_0 \sim 0.15$ kOe and saturation
is reached at $H_0 \sim 6.6$ kOe. Again the fifth and seventh
generations exhibit a nonzero magnetization in the low-field
region.

We have also studied cases where the applied field does not lie
along the easy axis ($\theta_H \neq \theta_{ua}$), and an example
for $N = 3$ is shown in Fig.\ 3. When the field is applied at
$45^\circ$ and $90^\circ$ from the easy axis, the magnetizations
continuously rotate towards the field direction. There is no
first-order phase transition in these configurations. The
saturation fields for $45^\circ$ and $90^\circ$ are $H_0 \sim 2.3$
kOe and $2.9$ kOe, respectively.

In conclusion, we have modelled the magnetization versus applied
field curves of Fe/Cr Fibonacci multilayers including both BL and
BQ exchange. The role of uniaxial anisotropy is studied, including
situations when the applied field direction is different from the
easy axis. One can see a discrete self-similar pattern in the
system when $H_{bq} << |H_{bl}|$ (Fig.\ 1) and most transitions
are of the first-order type. This property is highlighted by the
boxes drawn as insets in Fig.\ 1. Specifically, the magnetization
curves of higher generation $N$ reproduce some aspects of the
magnetization curves of lower generation $N-2$. By contrast, when
$H_{bq} \sim |H_{bl}|$ and most of the transitions are
second-order type, there is no apparent self-similarity. This
behaviour, regarding the effect of BQ exchange, differs from that
in quasiperiodic multilayers having cubic anisotropy \cite{4,5}.
We may surmise that the lower symmetry of uniaxial anisotropy
changes the nature of the phase transitions and consequently the
conditions for a self-similar pattern to occur in the
magnetization curves. An analogous behaviour may also apply for
the transport properties of the system and this provides a topic
for future studies.

We gratefully acknowledges a fellowship (to CGB) from the
Brazilian Research Council CNPq and partial financial support (to
MGC) from NSERC of Canada.

\newpage

\centerline {\bf Figure Captions}
\begin{enumerate}

\item Fig.\ 1. Magnetization (in units of the saturation magnetization)
versus applied field $H_0$ for (a) $N = 3$, (b) $N = 5$ and (c) $N
= 7$. The parameters are $H_{bl} = -1.0$ kOe, $H_{ua} = 0.5$ kOe,
$H_{bq} = 0.1$ kOe and $\theta_H = \theta_{ua} = 0^\circ$.

\vskip 0.5cm

\item Fig.\ 2. As in Fig. 1 but for $H_{bq} = 0.7$ kOe.

\vskip 0.5cm

\item Fig.\ 3. Magnetization (normalized) versus applied field $H_0$ for
three different values of angle $\theta_H$, taking $N = 3$,
$\theta_{ua} = 0^\circ$ and other parameters as in Fig.\ 1.

\end{enumerate}

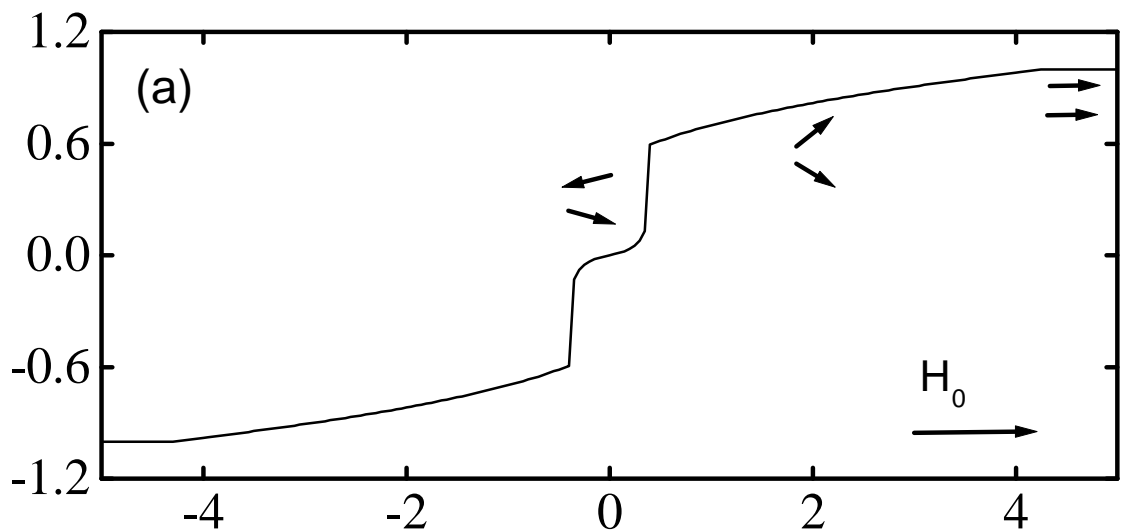
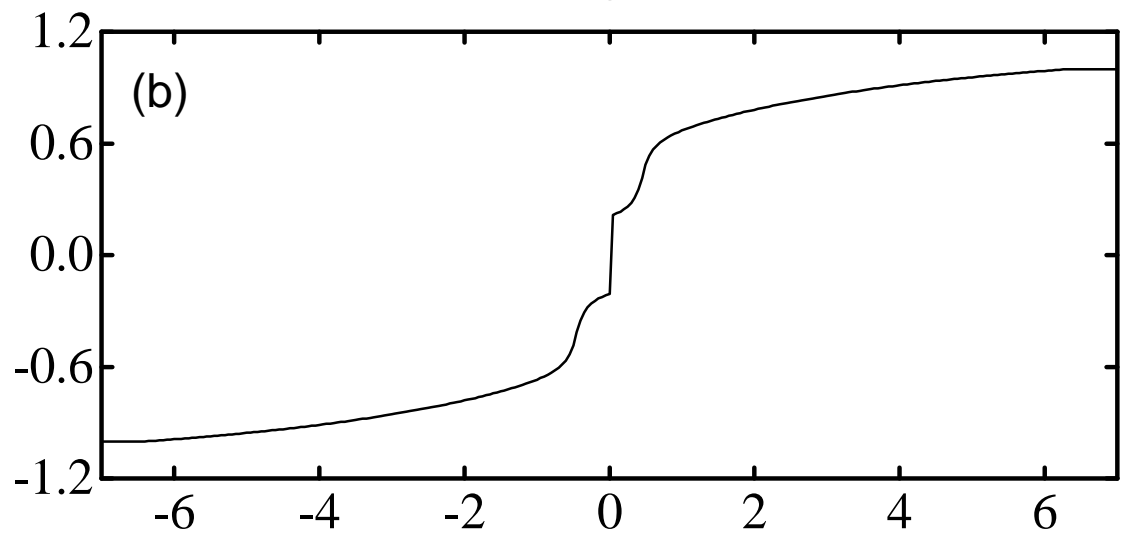
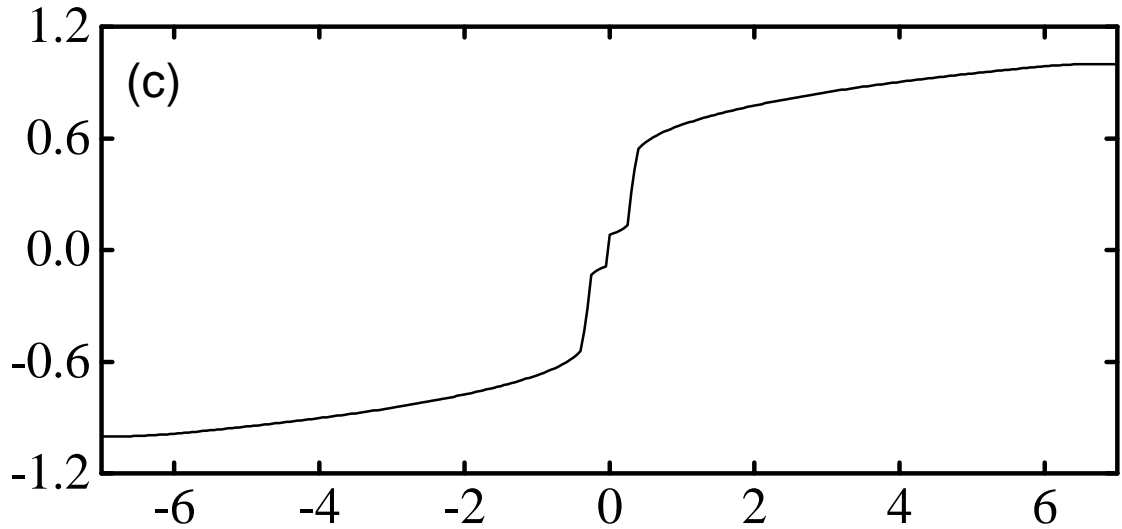
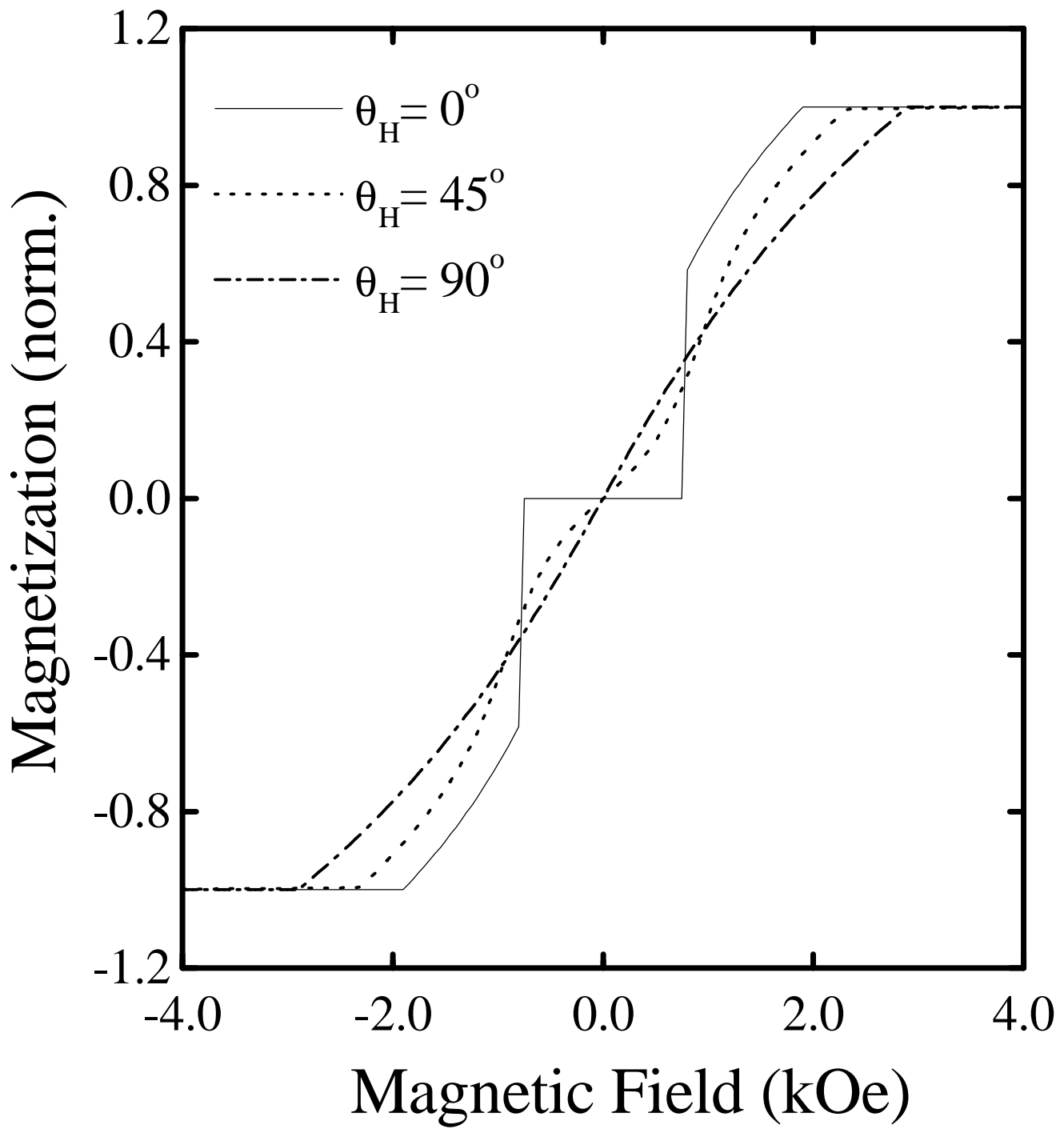
\end{document}